\documentclass[dvips,%
               review,%
               sort&compress,%
               ]{elsarticle}

\usepackage{amsmath}
\usepackage{amsfonts}
\usepackage{graphicx}
\usepackage{psfrag}
\usepackage{wasysym}
\usepackage{pinlabel}
\usepackage[latin1]{inputenc}
\usepackage{ae,aecompl} % instead of T1 fontenc allows scalable even
\usepackage[a4paper]{hyperref} % must come last!
\hypersetup{
    bookmarks=true,         % show bookmarks bar?
    unicode=false,          % non-Latin characters in Acrobat's bookmarks
    pdftoolbar=true,        % show Acrobat's toolbar?
    pdfmenubar=true,        % show Acrobat's menu?
    pdffitwindow=false,     % window fit to page when opened
    pdfstartview={FitH},    % fits the width of the page to the window
    pdftitle={My title},    % title
    pdfauthor={Tobias Schlüter},     % author
    pdfsubject={Sandwich Veto},   % subject of the document
    pdfnewwindow=true,      % links in new window
    colorlinks=false,       % false: boxed links; true: colored links
    linkcolor=red,          % color of internal links
    citecolor=green,        % color of links to bibliography
    filecolor=magenta,      % color of file links
    urlcolor=cyan           % color of external links
}

\newcommand{\mm}{\ensuremath{\textrm{mm}}}
\newcommand{\cm}{\ensuremath{\textrm{cm}}}

\newcommand{\MeV}{\ensuremath{\textrm{MeV}}}
\newcommand{\MeVc}{\ensuremath{\textrm{MeV}\!/\textrm{c}}}

\newcommand{\GeV}{\ensuremath{\textrm{GeV}}}
\newcommand{\GeVc}{\ensuremath{\textrm{GeV}\!/\textrm{c}}}

\title{Large-Area Sandwich Veto Detector with WLS Fibre Readout for
  Hadron Spectroscopy at COMPASS}
\author{T.~Schlüter\corref{cor}}\ead{tobias.schlueter@physik.uni-muenchen.de}
\author{W.~Dünnweber}
\author{K.~Dhibar\fnref{kd}}
\author{M.~Faessler}
\author{R.~Geyer}
\author{J.-F.~Rajotte\fnref{jf}}
\author{Z.~Roushan\fnref{zr}}
\author{H.~Wöhrmann}
\address{Department für Physik,
  Ludwig-Maximilians-Universität München,
  Am Coulombwall 1, D-85748 Garching, Germany}
\cortext[cor]{Corresponding author}
\fntext[kd]{Present address: Galgotias College of Engineering \& Technology,
  Greater Noida, India}
\fntext[jf]{Present address: MIT 26-441, 77 Massachusetts Ave.,
  Cambridge, MA 02139, USA}
\fntext[zr]{Present address: Klinikum Augsburg, Medizinische Physik
  und Strahlenschutz, 86156 Augsburg, Germany}

\begin{document}

\begin{abstract}
  A sandwich detector composed of scintillator and steel-covered lead
  layers was introduced in the fixed-target COMPASS experiment at CERN
  for vetoing events not completely covered by the two-stage magnetic
  spectrometer.  Wavelength shifting fibres glued into grooves in the
  scintillator tiles serve for fast read-out.  Minimum ionizing
  particles impinging on the $2\,\textrm{m} \times 2\,\textrm{m}$
  detector outside of a central hole, sparing the spectrometer's
  entry, are detected with a probability of 98\%.  The response to
  charged particles and photons is modeled in detail in Monte Carlo
  calculations.  Figures of merit of the veto trigger in $190\,\GeVc$
  $\pi^- + p$ (or nucleus) experiments are an enrichment of exclusive
  events in the recorded data by a factor of 3.5 and a false-veto
  probability of 1\%.
\end{abstract}

\maketitle
%\tableofcontents

\section{Introduction}
\label{sec:introduction}
COMPASS campaigns 2008 and 2009 were devoted mainly to light meson
spectroscopy via diffractive or central production.  A liquid hydrogen
target or, alternatively, a staggered Pb and W target, was bombarded
with $190\,\GeVc$ positive and negative hadron beams ($\pi^{\pm}$,
$K^{\pm}$, $p$) from the CERN M2 beamline.  Forward going particles
were detected in a two-stage magnetic spectrometer with tracking and
calorimetry in both stages~\cite{Abbon:2007pq}.

The requested event types have in common that the target proton
remains intact.  It is emitted at large angles compared to the beam
direction with momenta on the order of a few hundred $\MeVc$.  This
proton is detected by a recoil proton detector (RPD) which encloses
the target with an angular range from $55^{\circ}$ to $125^{\circ}$ as
seen from the centre and delivers a trigger signal on seeing a charged
particle track~\cite{Alexeev:2011}.  Apart from the recoiling proton
no other particle is expected in the target fragmentation region.
Inelastic, non-diffractive processes and diffractive excitations of
the target produce the dominant background, yielding additional
particles in the target fragmentation region.  Mostly these fall
outside the opening angle of the COMPASS spectrometer which amounts to
approx.~$\pm 10.3^{\circ}$ about the beam direction as seen from the
upstream target edge.  In order to enrich the recorded data with
kinematically complete events of the required type, a veto trigger
counter was needed that spans a large part of the angular range
between the acceptances of the spectrometer and that of the RPD.

\begin{figure}[tbp]
  \centering
  \psfrag{tagA}{\footnotesize a)}
  \psfrag{tagB}{\footnotesize b)}
  \psfrag{tagC}[lc][lc]{\footnotesize c)}
  \psfrag{tagD} {\footnotesize d)}
  \psfrag{tagE}[lc][lc]{\footnotesize e)}
  \psfrag{tagScale}[tc][tc]{\footnotesize $1\,\textrm{m}$}
  \includegraphics[width=.8\textwidth]{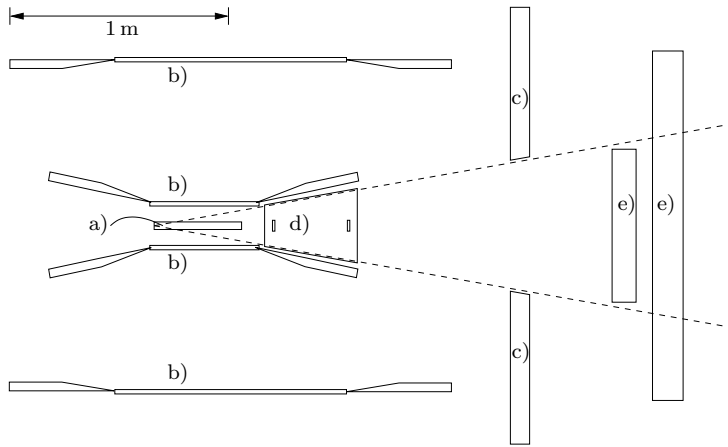}
  \caption{Schematic view of the spectrometer entry region of the
    COMPASS experiment showing the H2 target (a), inner and outer ring
    of the recoil proton detector with lightguides and
    photomultipliers (b), sandwich veto detector (c), cold silicon
    trackers in their conical cryostat (d), the first Micromegas and DC
    stations (e), and the vertical acceptance range of the
    spectrometer magnet ($10.3^{\circ}$, dashed line).}
  \label{fig:schematic}
\end{figure}

The sandwich detector (SW) covered in this report is a $2\,\textrm{m}
\times 2\,\textrm{m}$ detector with optically active wavelength
shifting (WLS) fibre readout that detects photons and charged
particles falling in a solid angle of $1.15\,\textrm{sr}$ outside of
the spectrometer's geometric acceptance (Fig.~\ref{fig:schematic}).
In principle the detector is a thin electromagnetic-calorimeter-type
detector complying with the following demands: high rate capability,
good time resolution, compact geometry, high efficiency for minimum
ionizing particles and photons with energy above $100\,\MeV$, low
false-veto probability.

In coincidence with the RPD trigger, a beam defining trigger and a
veto from a small counter for non-interacting beam particles, placed
at $33\,\textrm{m}$ downstream from the target, the SW veto
contributes to the hadron physics trigger.  At full use of the data
recording capacity of $30\,\textrm{kHz}$, it is found that inclusion
of the SW veto enriches the recorded sample with good candidate events
by a factor of about $3.5$.  Apart from the event type described
above, the SW veto is also useful in studies of exclusive events with
excited recoil protons detected by the RPD or with nuclear recoils
unnoticed by the RPD.

\section{Detector Setup}
\subsection{Scintillators with WLS fibre readout}

Light signals from scintillating material can be collected, converted
and transported to photomultipliers by wavelength shifting fibres.
This technique
% MF , implying wavelength shifting both in the scintillator and in
% the fibre,
has been optimized by the Moscow-Protvino
groups~\cite{Karyukhin1996415,Ivashkin1997321,Mineev2002362,Yershov2005454}
with regard to good efficiency, fast timing and radiation hardness.
It allows for compact large-area calorimeter designs.

The scintillator material used here is a polystyrene (DOW Styron 637)
with additives of pTP ($2\%$ weight fraction) as primary fluor and
POPOP ($0.02\%$ weight fraction) as secondary fluor.  The intensity of
the scintillation light assumes its maximum at $420\,\textrm{nm}$, the
refractive index is $n=1.59$.

Tiles of an area $20\,\textrm{cm}\times 20\,\textrm{cm}$ and thickness
of $5\,\textrm{mm}$ were manufactured at the the Institute of
High-Energy Physics (Protvino) by means of molding
techniques~\footnote{Details at
  \url{http://www.ihep.ru/scint/index-e.htm}}.  Eight grooves of
$1.4\,\textrm{mm}$ width and $2.2\,\textrm{mm}$ depth run in parallel
over each tile for fibre accomodation.  Four of these were included in
the mold, the other four milled subsequently.  Two small knobs were
molded and a central hole was drilled to allow for stable stacking in
sandwich layers.

Optically active fibres of $120\,\textrm{cm}$ length are used for
readout.  The chosen type of multiclad fast green wavelength shifters
(Bicron BCF-92, $\diameter 1\,\textrm{mm}$~\cite{StGobain:2005}) are
characterised by absorption and emission maxima at $410$ and
$490\,\textrm{nm}$, respectively.  Core and cladding refractive
indices are $n=1.60$ and $1.42$, respectively, and the $1/e$
attenuation length is $> 3.5\,\textrm{m}$.

\begin{figure}[tbp]
  \centering
  \includegraphics[width=.8\textwidth]{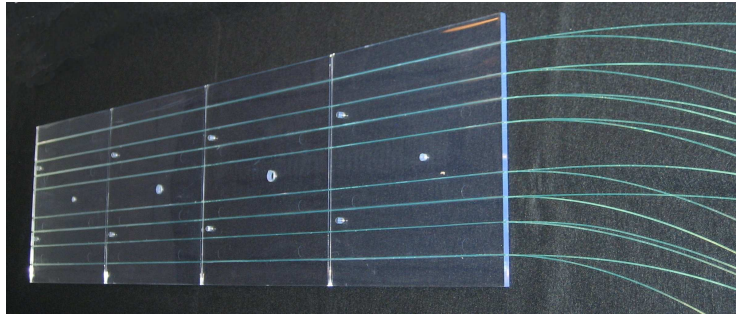}
  \caption{Single layer of 4 scintillator tiles with an area of
    $80\,\cm\times 20\,\cm$.  Eight pairs of fibres are glued into
    grooves of $2.2\,\mm$ depth.}
  \label{fig:layer}
\end{figure}
For installation in the sandwich detector, 4 scintillator tiles or 4
stacked pairs of tiles are grouped in bars of an area of $80\,\cm
\times 20\,\cm$ and thickness $5\,\mm$ or $10\,\mm$, respectively.  A
four-tile bar has eight pairs of fibres for common readout
(Fig.~\ref{fig:layer}).  Double layers consist of a layer as above,
the fibres running on the upper side, on which another layer is
stacked carrying only one fibre per groove, altogether $8\times 3$
fibres.

The fibres are glued into the grooves with optical cement
BC-600~\cite{StGobain:2002} which has a refractive index $n = 1.56$.
After mixing, this two-component glue is extruded into the groove with
a syringe.  The fibres are inserted such that they are fully covered
with glue.  For fibre-pair readout, a second string of glue is
overlayed on the first fibre before inserting the second fibre.  The
far-side ends as seen from the photomultiplier are cut at an angle to
suppress light reflection.  This is validated by the absence of
reflected photon signals.  After hardening of the optical cement,
scintillator bars have sufficient mechanical stability for subsequent
handling.

In the final assembly the scintillator bars are wrapped with a single
layer of white Tyvek paper R1025D (thickness $140\,\mu\textrm{m}$).
This diffusive reflector is found to increase the light output by
$25\%$.

Bundles of fibres from a single detector block (see below) are fed
through a plexiglas cylinder.  After fixation with epoxy glue, ends
are cut and the front side is polished.  It is attached to a XP2262B
phototube.  Initially XP2020 phototubes were employed, which have less
gain but elsewise are suited as well.  Both tubes are fast and provide
quantum efficiencies of about $20\%$.  The efficiency curves are
rather flat, peaking at $420\,\textrm{nm}$~\cite{Photonis:2007}.

\begin{figure}[tbp]
  \centering
  \psfrag{voltage}[br][br]{\footnotesize Voltage $\left[\textrm{mV}\right]$}
  \psfrag{time}[tr][tr]{\footnotesize Time $\left[\textrm{ns}\right]$}
  \includegraphics[width=.48\textwidth]{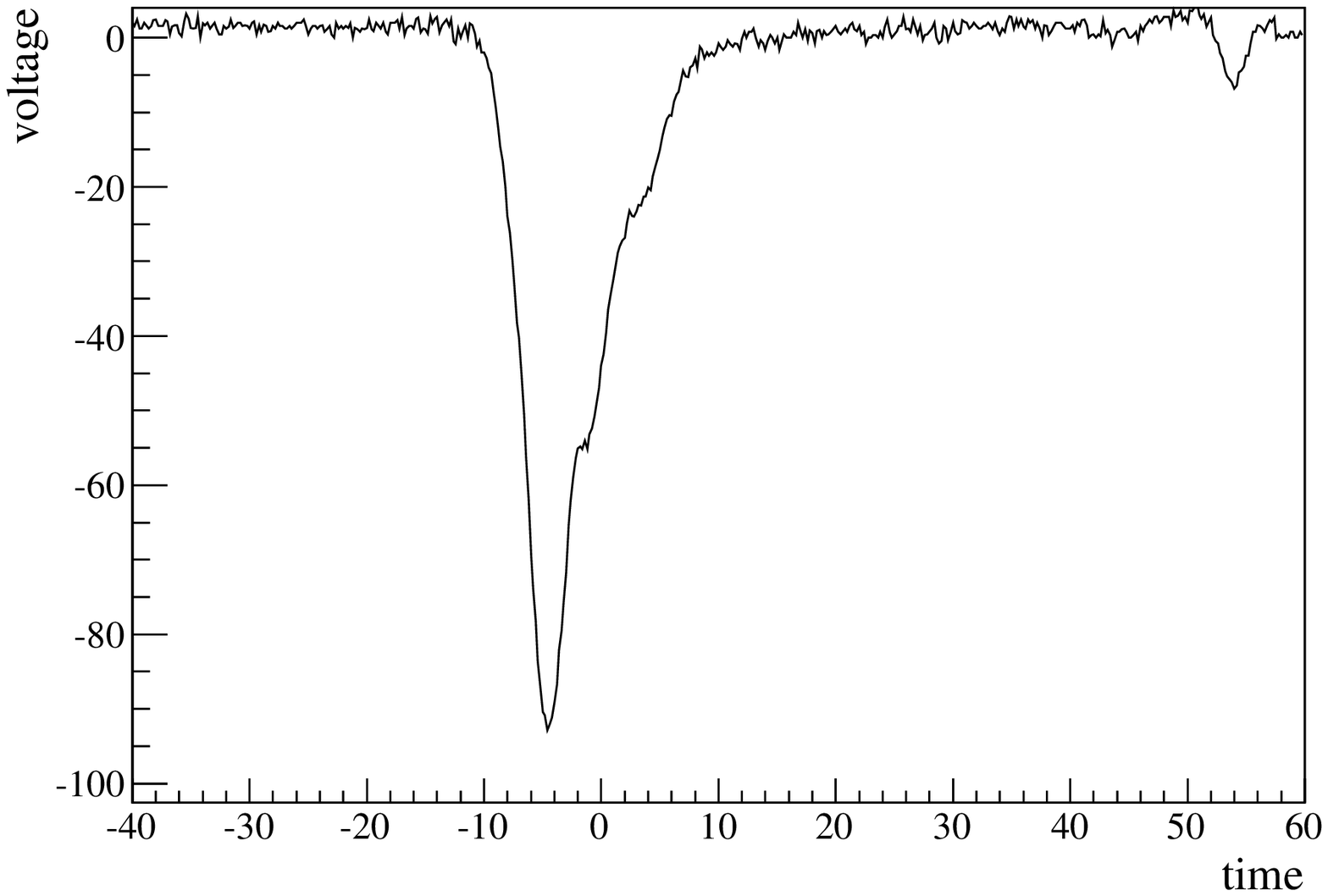}
  \includegraphics[width=.48\textwidth]{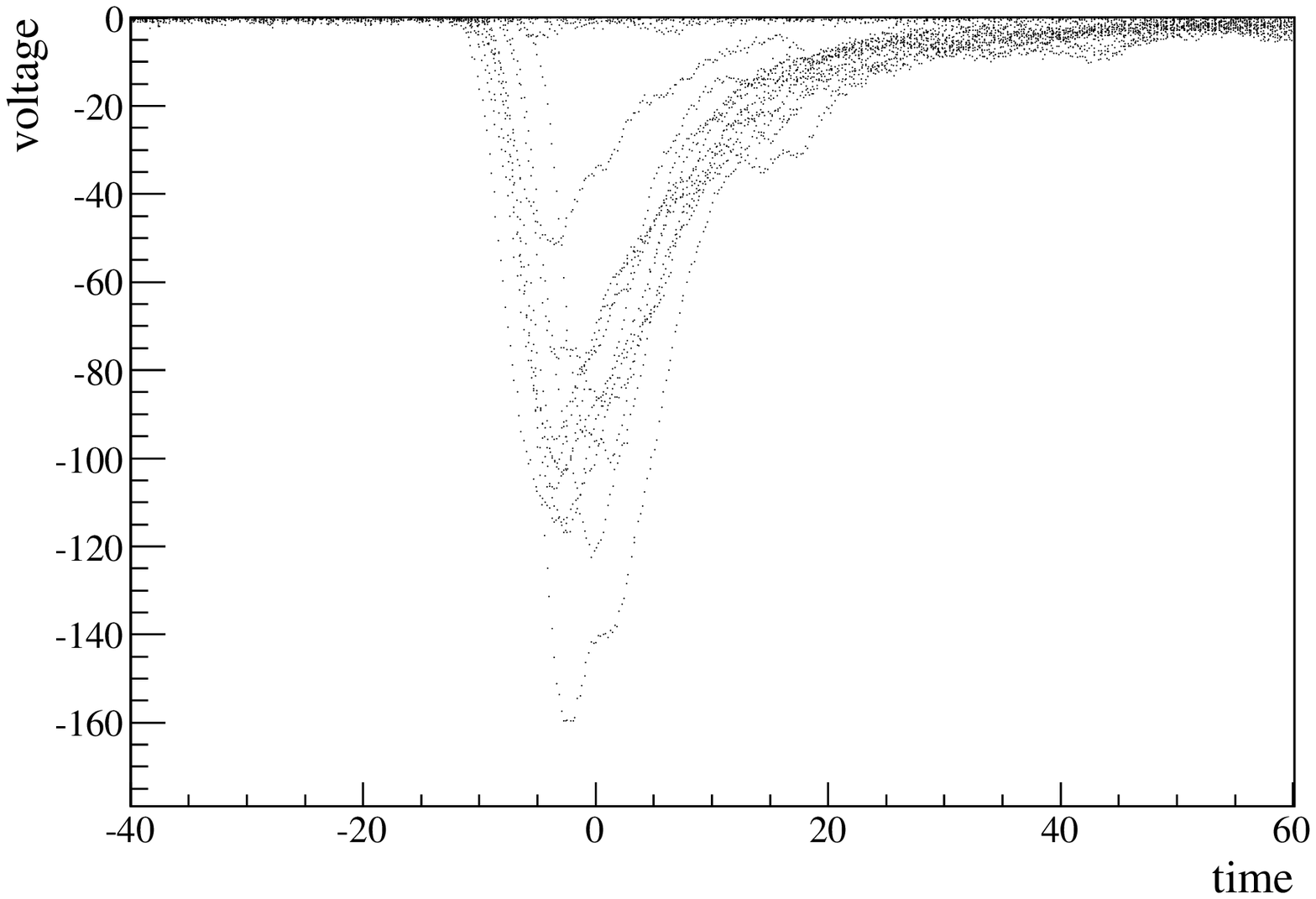}

  \caption{Pulses from cosmic muons recorded with an oscilloscope
    connected to the photomultiplier anode from one bare double layer
    of $10\,\mm$ scintillator thickness (left), and from the final
    sandwich detector (right).  An accidental single photoelectron
    show up on the tail of the left signal.}
  \label{fig:scope}
\end{figure}

Cosmic ray tests were performed with trigger counters above and below
scintillator tiles.  Pulses from muons in a double layer, with most
probable energy deposit of $1.9\,\MeV$, produce signals of about 10
photoelectrons (Fig.~\ref{fig:scope}).  Rise times and decay times
(20\%-80\%) of 3 and $6\,\textrm{ns}$ are found.  Comparing signals
from the far and near end, we find a light attenuation of 25\% over
the length of $80\,\cm$.

\subsection{Sandwich Assembly}

\begin{figure}[htbp]
  \centering
  \includegraphics[width=.9\textwidth]{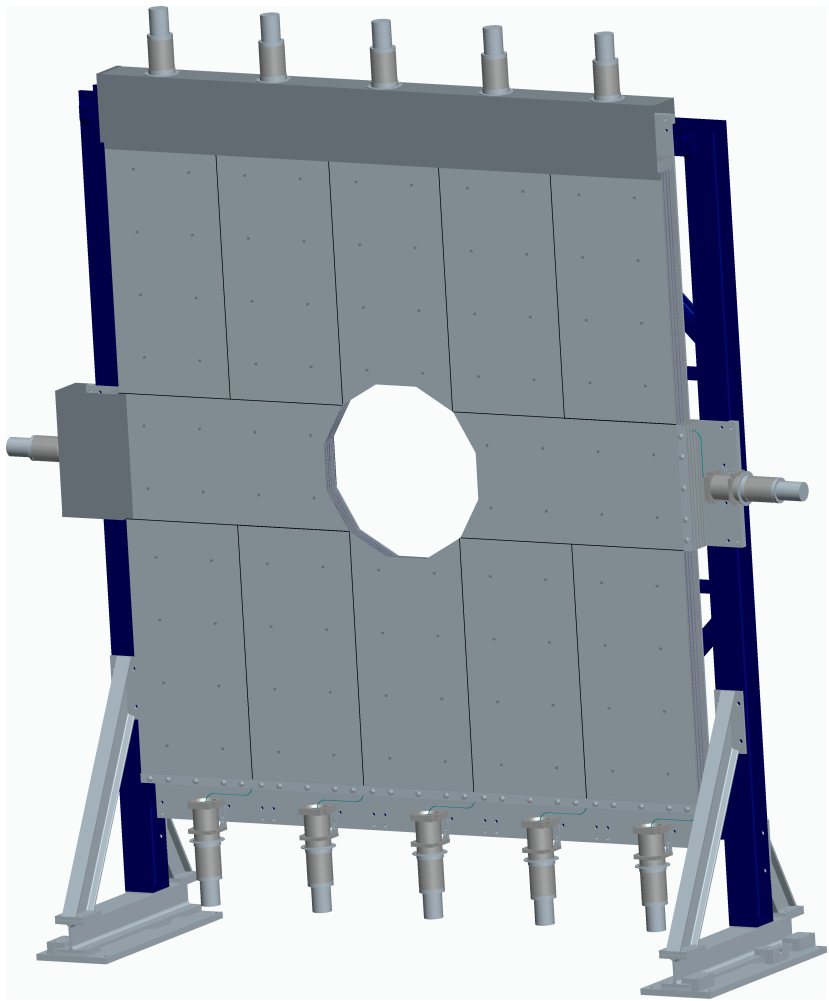}
  \caption{Design of the veto detector.  The light-shielding boxes are
    omitted for the lower 5 detector blocks and for one of the two
    horizontal detector blocks to uncover the photomultiplier
    assembly.  The support frame has horizontal and vertical lengths
    of 228 and $256\,\cm$, respectively.}
  \label{fig:assembly}
\end{figure}

The SW has a total thickness of 5.1 radiation lengths.  It consists of
12 blocks, each of an area of $80\,\cm\times 40\,\cm$ minus cutouts
for the central hole arranged on the $2\,\textrm{m}\times
2\,\textrm{m}$ surface (Fig.~\ref{fig:assembly}).  Each block is
assembled from 5 steel-covered lead-plates and 5 scintillator layers,
3 of $10\,\mm$ and 2 of $5\,\mm$ thickness.  As seen from the target
entry, the hole is approximately conical with an opening angle of $\pm
10.3^{\circ}$.  Each block is mounted on a steel plate of $8\,\mm$
thickness which is attached to an outer support frame.  The complete
detector has a mass of $2\,\textrm{t}$.  It is held in the vertical
position by a welded H-iron support frame which is stabilized by
struts across the corners.  These allow mounting the detector blocks
in the horizontal position.  The frame also carries a small
multiplicity counter covering the inner hole on the spectrometer
side~\cite{Alexeev:2011}.

\begin{figure}[hbtp]
  \labellist
  \small\hair 2pt
  \pinlabel {PMT} at 39 180
  \pinlabel {Detector Block} at 252 237
  \endlabellist
  \centering
  \includegraphics[width=.98\textwidth]{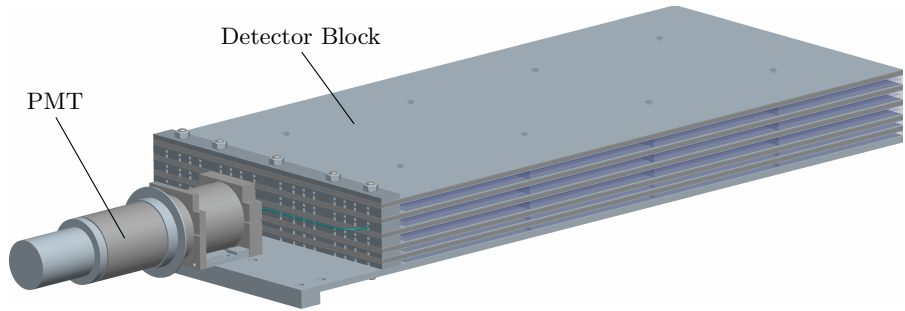}
  \caption{View of a detector block with photomultiplier tube (PMT)
    enclosed in a cylinder of soft iron which is attached to the lower
    steel plate.}
  \label{fig:module}
\end{figure}
\begin{figure}[htbp]
  \labellist
  \small\hair 2pt
  \pinlabel {Steel} [l] at 153 305
  \pinlabel {Lead} [l] at 292 305
  \pinlabel {Scintillator} [l] at 438 305
  \pinlabel {\rotatebox{90}{\scriptsize $10\,\textrm{mm}$}} [r] at 0 245
  \pinlabel {\rotatebox{90}{\scriptsize $5\,\textrm{mm}$}} [r] at 0 108
%  \pinlabel {Screw Head} [l] at 750 292
  \pinlabel {Al shims} [l] at 750 214
  \pinlabel {$20\,\textrm{cm}$} at 360 -16
  \endlabellist
  \centering
  \includegraphics[width=.9\textwidth]{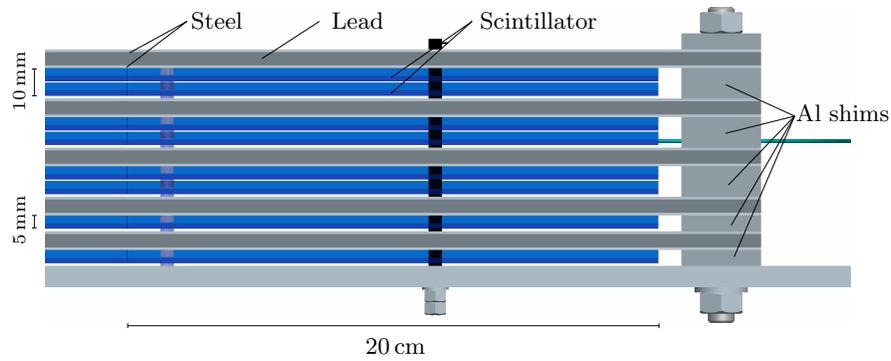}
  \caption{Lateral cut of the near (photomultiplier) side of a
    detector block.  The cut is along one of the grooves in the
    scintillator tiles.  One light fibre, out of a total 208, is shown
    emerging from a groove.  Steel screws cramp the 5 steel-covered
    lead layers and the Al shims to the lower steel plate which is
    downstream in the vertical detector position.  One of the bolts
    traversing the detector block is recognized.}
  \label{fig:module-close}
\end{figure}
The optically inactive layers of the calorimeter consist of $5\,\mm$
lead plates which have $1\,\mm$ steel glued to each side to accomplish
sufficient stiffness for the assembly (Figs.~\ref{fig:module},
\ref{fig:module-close}).  Each scintillator layer is formed by a pair
of $80\,\cm\times 20\,\cm$ scintillator bars lying side-by-side.
Their positions are fixed by two knobs per tile, mentioned above, and
by 8 bolts per detector block, traversing the central holes of the
scintillator tiles (Fig.~\ref{fig:module}, \ref{fig:module-close}),
four of which have sleeves between the lead/steel plates ensuring
their proper distance.  To accomplish mechanical stability without
exerting pressure on the scintillators, the lead/steel plates are
cramped onto the rear steel plate where the correct distance is
enforced by aluminium shims.  In total a block has 64 scintillator
plates which are grouped in 3 double layers on the upstream side (top
in Figs.~\ref{fig:module}, \ref{fig:module-close}) and 2 single layers
on the downstream side.  For the given amount of scintillators this
order gives the best efficiency for low-energy photons according to
Monte Carlo simulations (see below).

The light fibres are fed through channels in the shims.  For common
readout, the total of 208 fibres per block are bundled in a plastic
cylinder which is attached to the entrance window of a XP2262B
phototube.  The stray magnetic field of the spectrometer's first stage
dipole magnet necessitates magnetic shielding of the photomultipliers.
This is accomplished by double cylinders made of $\mu$-metal and soft
iron, respectively (Fig.~\ref{fig:module}).

\subsection{Monte Carlo Studies}

\begin{figure}[tbp]
  \centering
  \psfrag{Edep}[tr][tr]{\footnotesize $\Delta E \left[\MeV\right]$}
  \psfrag{Intensity}[br][br]{\footnotesize Intensity (a.u.),
    Efficiency}
  \psfrag{gamma50}{\footnotesize $\gamma$, $E = 50\,\MeV$}
  \psfrag{gamma100}{\footnotesize $\gamma$, $E = 100\,\MeV$}
  \psfrag{gamma200}{\footnotesize $\gamma$, $E = 200\,\MeV$}
  \psfrag{mip}{\tiny A) $\mu^{-}$, $E_{\textrm{kin}} =
    285\,\MeV$}
  \psfrag{tagA}{\tiny A}
  \psfrag{mu160}{\tiny B) $\mu^-$, $E_{\textrm{kin}} = 160\,\GeV$}
  \psfrag{tagB}{\tiny B}
  \includegraphics[width=.9\textwidth]{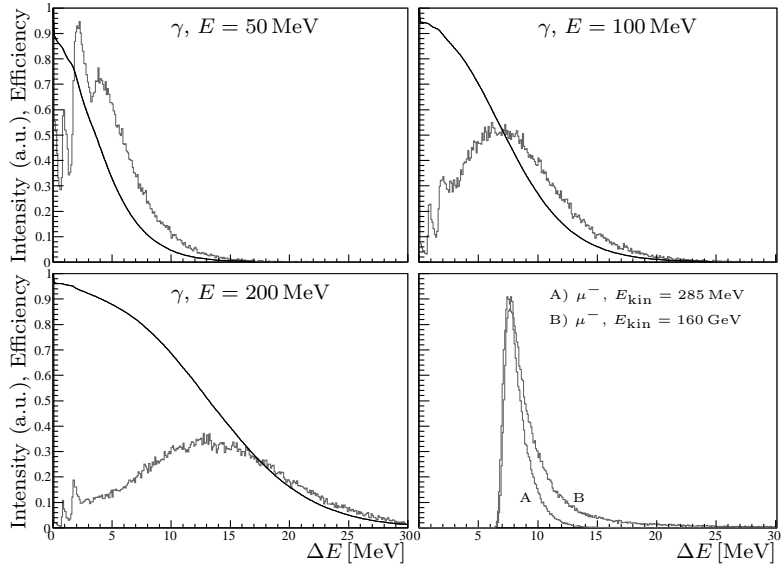}
  \caption{Monte Carlo prediction of the energy deposit, summed over
    all scintillator layers, for photons and
    muons impinging perpendicularly on the veto detector (grey
    histograms).  For photons, the black lines indicate the fraction
    of events above the value of $\Delta E$, thus giving the
    efficiency as function of thresholds on the energy deposit.}
  \label{fig:mc_photon}
\end{figure}

The response of the detector to photons and charged particles was
simulated with the Monte Carlo code {\sc
  Geant4}~\cite{Agostinelli:2002hh,Allison:2006ve}.\footnote{All plots
  were created using the {\tt QGSP\_BIC} physics list.  No significant
  dependency on the choice of physics list was observed.}  The energy
deposit in the scintillator layers for perpendicularly impinging
photons of various energies (Fig.~\ref{fig:mc_photon}) is
characterized by a broad peak from showers and narrow peaks
superimposed at lower energies.  The first (second) narrow peaks are
due to single electrons or, less frequently, positrons traversing only
one of the scintillator layers of $5\,\mm$ ($10\,\mm$) thickness,
respectively.  Shown for comparison is the energy deposit of muons
with $E_{kin}=285\,\MeV$ and $E_{kin}=160\,\GeV$ (used at COMPASS for
spectrometer alignment).  The most probable energy deposit of muons is
almost independent of their energy in this range.  This comes in spite
of the energy loss at $160\,\GeV$ being considerably larger than in
the minimum ionizing case ($285\,\MeV$) because of the occurence of
radiative processes: the ensuing photons and $e^+e^-$ pairs are mostly
produced and stopped in a lead layer depositing no energy in the
scintillators.

The fraction of photons with energy deposit above $\Delta E$ gives the
efficiency for discriminator threshold corresponding to $\Delta E$
(Fig.~\ref{fig:mc_photon}).  For threshold corresponding to one third
of $\Delta E_{\textrm{MIP}}=7.5\,\MeV$, the most probable energy
deposit of minimum ionizing particles, the efficiency is above $90\%$
for energies above $100\,\MeV$ and drops to $80\%$ for $50\,\MeV$
photons.  This threshold was chosen in COMPASS runs after a pulse
height calibration with muons of high energy (next section), however,
halving the threshold is feasible at the noise level observed in the
experiment.

In order to suppress delayed energy deposits from secondary processes
induced by very soft hadrons, the simulated energy deposits were
integrated over a timespan of $10\,\textrm{ns}$.  This timespan is
adapted to the observed signal broadening coming from light collection
and conversion in the scintillators and the WLS fibres (see below).
Significant differences from the complete energy deposits occur only
for hadrons of low energy.

% The MC simulation does not take light collection into account where
% most of the time spread ($O(10\,\textrm{ns})$) happens.  Simulated
% energy deposits were not taken into account if they happened more than
% $10\,\textrm{ns}$ after the start of the simulated event.  However,
% significant differences from the complete energy deposits occur only
% for hadrons of low energy.

% The actual pulse height which is relevant for the electronic
% discrimination depends not only on the energy deposit but also on the
% timing properties of the system.  Time spreading occurs in the
% interactions in the calorimeter and also in the scintillation and
% light collection processes.  The effect of the latter processes
% (disregarded in the present MC simulations) is included in a pulse
% height calibration with muons of high energy (next section).  In this
% way the relation between threshold height and energy deposit is
% established for photons and minimum ionizing particles. As borne out
% by the MC simulations, in these cases the energy deposit in the
% calorimeter is much faster than the secondary light collection which
% requires about $10\,\textrm{ns}$.  In the present calculations of
% energy deposits, an integration time of $10\,\textrm{ns}$ was
% chosen. However, significant differences from the complete energy
% deposits occur only for hadrons of low energy.

\begin{figure}[tbp]
  \centering
  \psfrag{Edep}[tr][tr]{\footnotesize $\Delta E \left[\MeV\right]$}
  \psfrag{Intensity}[br][br]{\footnotesize Intensity (a.u.), Efficiency}
  \psfrag{pi50}{\footnotesize $\pi^-$, $E_{\textrm{kin}} = 50\,\MeV$}
  \psfrag{pi100}{\footnotesize $\pi^-$, $E_{\textrm{kin}} = 100\,\MeV$}
  \psfrag{pi200}{\footnotesize $\pi^-$, $E_{\textrm{kin}} = 200\,\MeV$}
  \psfrag{pipl50}{\footnotesize $\pi^+$, $E_{\textrm{kin}} = 50\,\MeV$}
  \includegraphics[width=.9\textwidth]{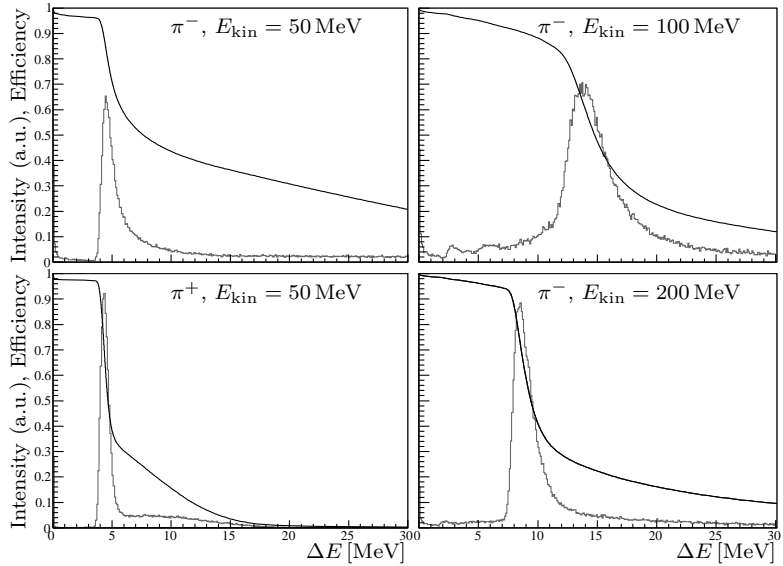}
  \caption{Monte Carlo prediction of the energy deposit, summed over
    all scintillator layers, for $\pi+$ and $\pi^-$ impinging
    perpendicularly on the veto detector (grey histograms).
    Integration of $\Delta E$ was stopped after $10\,\textrm{ns}$
    which is relevant only in the case of $50\,\MeV$ (see text).  For
    the black lines see Fig.~\ref{fig:mc_photon}.}
  \label{fig:mc_pi}
\end{figure}

For pions with kinetic energy above $50\,\MeV$, efficiencies above
$95\%$ are obtained (Fig.~\ref{fig:mc_pi}).  The edge at $4\,\MeV$ of
the distribution of energy deposits for $50\,\MeV$ pions is due to
their stopping in the second lead/steel layer.  The integration time
of $10\,\textrm{ns}$ is relevant for $50\,\MeV$ pions.  For larger
time spans the $\Delta E$ distribution develops a broad shoulder
towards larger values (not shown).  In the case of stopped negative
pions, this is mostly due to secondary neutrons scattering in the
scintillator material after being produced in the Pb layers.  In the
case of stopped positive pions, the broadening can be traced back to
$\mu^+(e^+\bar\nu_{\mu}\nu_e)$ decay.  These effects are not important
for the present vetoing at sufficiently low threshold but should be
kept in mind when pile-up is a concern.  Pions with energy at or above
$100\,\MeV$ yield a peak above or at $\Delta E_{\textrm{MIP}}$
(Fig.~\ref{fig:mc_pi}, right) with insignificant dependence on the
integration time and insignificant $\pi^+$/$\pi^-$ difference, since
the pions traverse the detector.

%   The
% small portion with lower energy deposit is attributed to hadronic
% interactions in the first lead/steel layer.  Negatively and positively
% charged pions stopped in the detector display a difference (left two
% plots) which is interesting though not of primary importance for
% threshold vetoing at sufficiently low threshold.  A significant
% fraction of stopped $\pi^+$ end up in the
% $\mu^+(e^+\nu_e\bar\nu_{\mu})\nu_{\mu}$ decay chain, where the
% neutrinos lead to a minimum in the energy deposit distribution (dashed
% grey line).  In contrast, low energy $\pi^-$ interact hadronically
% yielding nuclear fragments which contribute to the energy deposit.

The first lead/steel layer corresponds to the range of $30\,\MeV$
pions.  Up to this energy pions are not vetoed.  This layer was chosen
for the entry to suppress false vetoes from delta electrons
accompanying valid events in the target.  Monte Carlo simulations of
delta electron production by $190\,\GeV$ pions traversing the
$40\,\cm$ liquid hydrogen target give a $1\%$ probability for a veto
signal induced by delta electrons.

\section{Performance of the Veto Detector}
\label{sec:perf-veto-detect}

The SW detector was integrated into the hardware trigger in the 2008
and 2009 COMPASS campaigns with pion, proton and short-time muon runs.
The so-called physics trigger incorporated in addition the recoil
proton and beam triggers, see sec.~\ref{sec:introduction}.

Photomultiplier signals from cosmic muons traversing the final
detector assembly have average rise times (20\%-80\%) of
$3.3\,\textrm{ns}$, logarithmic decay times of $9\,\textrm{ns}$, and
widths at half maximum of $12\,\textrm{ns}$ (Fig.~\ref{fig:scope}).
These values are larger by a factor of almost 3 than those obtained
for single photons and they are also larger than the corresponding
values for a single double layer.  These differences are attributed to
the statistical spread of light collection times.

The total number of photoelectrons for minimum ionizing particles
(MIP) impinging perpendicularly on the detector amounts to 45.  Taking
the quantum efficiency into account, this corresponds to 220 photons
entering the photomultiplier window.  More relevant for fast
discrimination is the signal height.  Because of the photoelectrons'
time spread the resulting signal height corresponds to a lower number
of single photoelectrons.  It is found that the average signal height
for MIPs corresponds to 18.7 single photoelectrons.  For the COMPASS
veto trigger a discriminator threshold corresponding to six times the
single photoelectron level was chosen.  The $10\,\textrm{ns}$
coincidence time requirement of the COMPASS veto system was easily
fulfilled.

The veto efficiency for MIPs was determined with $160\,\GeVc$ muon
beams using a halo trigger.  A veto flag probability of 98\% was
obtained for muons with a reconstructed track traversing the SW
detector.  This value refers to homogeneous irradiation of the
complete detector plane excluding the central hole.  Tracks at the
block edges contribute more than half of the missing 2\%.

%\begin{figure}[hbtp]
%  \centering
%  \includegraphics[width=.8\textwidth]{p_vs_th.eps}
%  \caption{(Color online.) Scattering angle vs.\ momentum for events
%    with a single reconstructed track in the spectrometer.  The right
%   image shows events where no veto signal was produced by the
%     sandwich detector, whereas in the left picture there was such a
%     veto signal.  In the right picture the elastic scattering peak is
%     clearly visible at the nominal beam momentum of $191\,\GeV$, a
%     minimum scattering angle is enforced by the trigger condition that
%     there be a recoiling proton.  With the veto signal, the
%     distribution is widenend and shifted to lower momenta.}
%   \label{fig:pVsTh}
% \end{figure}

\begin{figure}[htbp]
  \centering
  \psfrag{thetamRad}[br][br]{\footnotesize Angle $\left[\textrm{mrad}\right]$}
  \psfrag{pGeV}[tr][tr]{\footnotesize $p \left[\GeVc\right]$}
  \includegraphics[width=.8\textwidth]{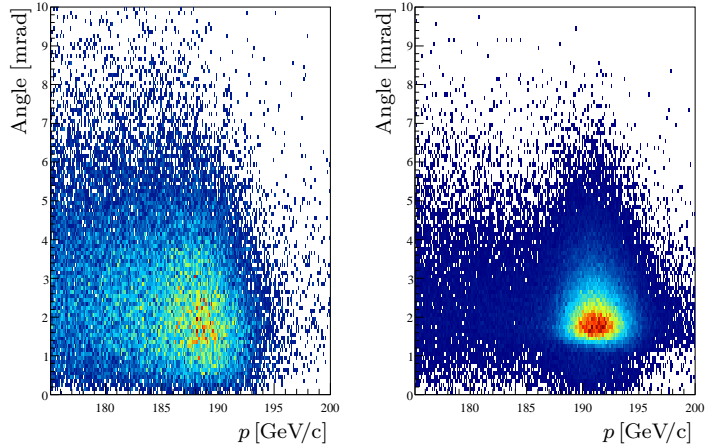}
  \caption{(Color online.)  Scattering angle vs.\ momentum for events
    from $191\,\GeVc$ $\pi^{-}+\textrm{LH}_2$ with three reconstructed
    charged tracks in the spectrometer.  The scattering angle and
    momentum $p$ refer to the total momentum vector in the laboratory
    sytem evaluated from the three tracks.  The left image shows
    events where a veto signal was produced by the sandwich detector,
    whereas in the right picture there was no such veto signal.  Small
    scattering angles are suppressed by the requirement of a signal
    from the recoil proton detector.}
  \label{fig:pVsTh3}
\end{figure}

% \begin{figure}[htbp]
%   \centering
%   \includegraphics[width=.8\textwidth]{delta_phi.eps}
%   \caption{Coplanarity angle for events with one reconstructed track.}
%   \label{fig:coplanarity}
% \end{figure}

\begin{figure}[hbtp]
  \centering
  \psfrag{Entries}[br][br]{\footnotesize Entries}
  \psfrag{DeltaphiRad}[tr][tr]{\footnotesize $\Delta\phi \left[\textrm{rad}\right]$}
  \includegraphics[width=.8\textwidth]{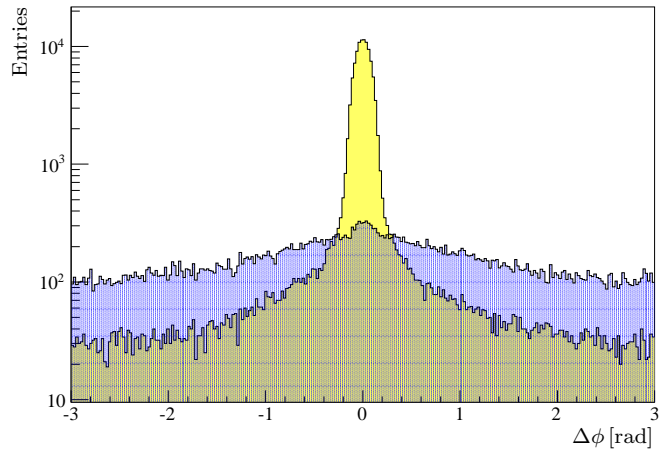}
  \caption{(Color online.)  Coplanarity angle (see text) for events
    with three reconstructed charged tracks with total momentum above
    $180\,\GeVc$ recorded with (patterned filling) / without veto
    signal from the sandwich detector.}
  \label{fig:coplanarity3}
\end{figure}
Hadron beam test runs with the SW excluded from the hardware trigger
but recorded as flag demonstrate its proper operation.  For events
carrying the flag, the total momentum distribution of particles
detected in the spectrometer is shifted towards lower momenta with
respect to the incoming $\pi^-$ momentum of $191\,\GeVc$
(Fig.~\ref{fig:pVsTh3} left) due to the momentum carried away into the
sandwich veto detector.  This is attributed to target fragmentation.
Such kind of events are strongly suppressed in the distribution
without veto flag where a peak characteristic of exclusive kinematics
stands out (Fig.~\ref{fig:pVsTh3} right).  This group of events is
attributed to the processes of interest here, namely peripheral
production of mesons.  For this example, events with three
reconstructed tracks in the spectrometer were selected, defining a
total momentum vector.  This vector and the momentum vector
$\boldsymbol{p}_i$ of the incoming pion define an (oriented) plane
which can be compared with the (oriented) plane defined by
$\boldsymbol{p}_i$ and the momentum vector of the target-like recoil
detected with the recoil detector.  The ``coplanarity angle''
$\Delta\phi$ between these planes shows a peak at zero, characteristic
of exclusive kinematics, for events with no SW veto signal, whereas
events with a SW veto signal have a broad $\Delta\phi$
distribution~(Fig.~\ref{fig:coplanarity3}).  A peak in the latter
distribution would indicate false vetoes.  For these plots a total
reconstructed momentum $> 180\,\GeVc$ was required, which largely
suppresses events where momentum was carried away by neutrals not
taken into account here.

Important figures of merit are the enrichment factor of useful data in
the recorded events and the probability of false vetos.  The former is
defined as the factor by which the physics trigger rate increases if
the sandwich is not included in the veto.  At given data recording
capability of about $30\,\textrm{kHz}$ in the present case, the rate
of physics data recording is increased by this factor due to the SW
veto.  Enrichment factors of 3.3 to 3.8 were obtained in the hadron
runs.  Typical beam rates were $2\cdot10^8$ particles per spill with
spill length and repetitions rate of 16.8 and $21\,\textrm{s}$,
respectively, and SW rate of about $1\,\textrm{MHz}$.  The probability
for wrong vetoes was extracted from SW coincidences with
noninteracting beam particles and also from the $\Delta\phi$
distributions as in Fig.~\ref{fig:coplanarity3}.  A probability of 1\%
was found for faulty SW vetoes, which agrees with estimates from the
simulation of delta electron production.

\section{Summary}

A compact $2\,\textrm{m}\times2\,\textrm{m}$ sandwich detector of 5.1
radiation lengths total thickness was installed around the COMPASS
spectrometer's entry for vetoing on incomplete detector events.
Read-out by WLS fibres running in grooves over the scintillators
allowed for MHz rates and fast triggering.  A total energy deposit in
the scintillators of $7.5\,\MeV$, obtained for minimum ionizing
particles (MIP), results in 220 photons entering the photomultipliers
(PM).  Statistical spread of light collection times leads to a signal
rise time of $3.3\,\textrm{ns}$ and logarithmic decay time of
$9\,\textrm{ns}$, corresponding to 3 times the single photo electron
values. The resulting MIP signal height corresponds to 18.7 photo
electrons. The PM output was fed directly into the veto discriminator.

The veto efficiency for MIP's was found to amount to $98\,\%$ for a
discriminator threshold corresponding to one third of the MIP signal
height. Monte Carlo simulations yield efficiencies for this threshold
which are above $95\,\%$ for pions with kinetic energy above 50 MeV
and above $90\,\%$ ($80\,\%$) for $100\,\MeV$ ($50\,\MeV$)
photons. The detector performed well for about 12 months of running
time in 2 years, mostly with pion beams of $190\,\GeVc$ impinging on a
liquid hydrogen target.  Rejecting events not completely covered by
the spectrometer, it increased the fraction of useful data with
complete kinematics by a factor of 3.5 as compared to data recording
with inactive sandwich veto.  Since this factor was obtained at full
use of the data recording capacity of $30\,\textrm{kHz}$, it
translates directly into a gain in statistics for the diffractive and
central production processes under study.

\section*{Acknowledgements}
We are grateful for the skillful design engineering by P. Hartung (MLL
Gar\-ching), for valuable advice from V.A. Polyakov, and for
assistance by H. Yang.  We also acknowledge financial support by BMBF
and by the DFG cluster of excellence ``Origin and Structure of the
Universe''.

\bibliography{general,photons}

\begin{thebibliography}{10}
\expandafter\ifx\csname url\endcsname\relax
  \def\url#1{\texttt{#1}}\fi
\expandafter\ifx\csname urlprefix\endcsname\relax\def\urlprefix{URL }\fi
\expandafter\ifx\csname href\endcsname\relax
  \def\href#1#2{#2} \def\path#1{#1}\fi

\bibitem{Abbon:2007pq}
P.~Abbon, et~al., {The COMPASS Experiment at CERN}, Nucl. Instrum. Meth. A577
  (2007) 455--518.
\newblock \href {http://arxiv.org/abs/hep-ex/0703049}
  {\path{arXiv:hep-ex/0703049}}, \href
  {http://dx.doi.org/10.1016/j.nima.2007.03.026}
  {\path{doi:10.1016/j.nima.2007.03.026}}.

\bibitem{Alexeev:2011}
M.~Alekseev, et~al., {The COMPASS 2008 Spectrometer}, to be submitted to Nucl.\
  Inst.\ Meth.\ A (2011).

\bibitem{Karyukhin1996415}
A.~Karyukhin, et~al., Radiation hardness study on molded scintillation tiles
  and wavelength shifting fibers, Nucl. Instrum. Meth. B117~(4) (1996) 415 --
  420.
\newblock \href {http://dx.doi.org/DOI: 10.1016/0168-583X(96)00301-1}
  {\path{doi:DOI: 10.1016/0168-583X(96)00301-1}}.

\bibitem{Ivashkin1997321}
A.~P. Ivashkin, et~al., {Scintillation ring hodoscope with WLS fiber readout},
  Nucl. Instrum. Meth. A394~(3) (1997) 321 -- 331.
\newblock \href {http://dx.doi.org/DOI: 10.1016/S0168-9002(97)00657-8}
  {\path{doi:DOI: 10.1016/S0168-9002(97)00657-8}}.

\bibitem{Mineev2002362}
O.~Mineev, et~al., {Photon sandwich detectors with WLS fiber readout}, Nucl.
  Instrum. Meth. A494~(1-3) (2002) 362 -- 368.
\newblock \href {http://dx.doi.org/DOI: 10.1016/S0168-9002(02)01493-6}
  {\path{doi:DOI: 10.1016/S0168-9002(02)01493-6}}.

\bibitem{Yershov2005454}
N.~Yershov, et~al., Long sandwich modules for photon veto detectors, Nucl.
  Instrum. Meth. A543~(2-3) (2005) 454 -- 462.
\newblock \href {http://dx.doi.org/DOI: 10.1016/j.nima.2004.11.051}
  {\path{doi:DOI: 10.1016/j.nima.2004.11.051}}.

\bibitem{StGobain:2005}
Saint-Gobain Inc., Data Sheet 03-05 (2005).

\bibitem{StGobain:2002}
Saint-Gobain Inc., Data Sheet 06-02 (2002).

\bibitem{Photonis:2007}
Photonis, Photomultiplier Tubes Catalogue (2007).

\bibitem{Agostinelli:2002hh}
S.~Agostinelli, et~al., {{\sc Geant} 4 -- a simulation toolkit}, Nucl. Instrum.
  Meth. A506 (2003) 250--303.
\newblock \href {http://dx.doi.org/10.1016/S0168-9002(03)01368-8}
  {\path{doi:10.1016/S0168-9002(03)01368-8}}.

\bibitem{Allison:2006ve}
J.~Allison, et~al., {Geant4 developments and applications}, IEEE Trans. Nucl.
  Sci. 53 (2006) 270.
\newblock \href {http://dx.doi.org/10.1109/TNS.2006.869826}
  {\path{doi:10.1109/TNS.2006.869826}}.

\end{thebibliography}

\end{document}